%% file: template.tex
\documentclass{article}

\usepackage{arxiv}

\usepackage[utf8]{inputenc} 
\usepackage[T1]{fontenc}    
\usepackage{hyperref}       
\usepackage{url}            
\usepackage{booktabs}       
\usepackage{amsfonts}       
\usepackage{nicefrac}       
\usepackage{microtype}      
\usepackage{lipsum}
\usepackage{graphicx}

\usepackage{times}
\usepackage{epsfig}
\usepackage{graphicx}
\usepackage{amsmath}
\usepackage{amssymb}
\usepackage{booktabs}
\usepackage{tabu}
\usepackage{tabularx}
\usepackage{xcolor}
\usepackage{graphicx}
\usepackage{stix}
\usepackage{subfigure}
\usepackage{fancyhdr}
\usepackage{mathtools}
\usepackage{mwe}
\usepackage{mathtools, cuted}
\usepackage{adjustbox}
\usepackage{multirow}
\usepackage[parfill]{parskip}
\usepackage{gensymb }

\title{ALBRT: Cellular Composition Prediction in Routine Histology Images}

\author{Muhammad Dawood\textsuperscript{1}, Kim Branson\textsuperscript{2}, Nasir M. Rajpoot\textsuperscript{1}, Fayyaz ul Amir Afsar Minhas\textsuperscript{1}\\ 
\textsuperscript{1}Department of Computer Science University of Warwick, Coventry, UK \\
\textsuperscript{2} Artificial Intelligence \& Machine Learning, GlaxoSmithKline \\
{\tt\small\{Muhammad.Dawood, N.M.Rajpoot, Fayyaz.Minhas\}@warwick.ac.uk, kim.m.branson@gsk.com} 

}

\begin{document}
\maketitle
\begin{abstract}
Cellular composition prediction, i.e., predicting the presence and counts of different types of cells in the tumor microenvironment from a digitized image of a Hematoxylin and Eosin (H\&E) stained tissue section can be used for various tasks in computational pathology such as the analysis of cellular topology and interactions, subtype prediction, survival analysis, etc. In this work, we propose an image-based cellular composition predictor (ALBRT) which can accurately predict the presence and counts of different types of cells in a given image patch. ALBRT, by its contrastive-learning inspired design, learns a compact and rotation-invariant feature representation that is then used for cellular composition prediction of different cell types. It offers significant improvement over existing state-of-the-art approaches for cell classification and counting. The patch-level feature representation learned by ALBRT is transferrable for cellular composition analysis over novel datasets and can also be utilized for downstream prediction tasks in CPath as well. The code and the inference webserver for the proposed method are available at the URL: \href{https://github.com/engrodawood/ALBRT}{https://github.com/engrodawood/ALBRT}.
\end{abstract}

\section{Introduction}

Predicting the presence of different types of cells and quantifying their counts in digitized images of Hematoxylin and Eosin (H\&E) stained tissue slides can be useful for various downstream prediction tasks in computational pathology (CPath) such as  survival prediction \cite{lu2018nuclear,shaban2019novel,ko2021body}, prognosis  \cite{yuan2021neutrophil,mantovani2017tumour}, recurrence prediction \cite{ji2019nuclear},  gene expression and biological process analysis \cite{he2020integrating, zhan2019correlation}. Cellular composition can give insights into cellular diversity in the tumor microenvironment (TME) \cite{galli2020relevance} as well as tissue organization in the whole slide image (WSI) which can be helpful in therapeutic decision making \cite{lin2019tumor}. For instance, a high proportion of stromal tumor-infiltrating lymphocytes (sTILs) are associated with good prognosis \cite{kuroda2021tumor,denkert2018tumour,dieci2019association}, whereas excessive tumor-associated macrophages infiltration signifies poor prognosis in breast, bladder, and cervix carcinomas \cite{bingle2002role,leek1996association}. Similarly, tumors with co-presence of immune cells and tumor cells in the TME can be considered potential candidates for immunotherapy \cite{abduljabbar2020geospatial,galli2020relevance}. Despite its usefulness, it is not feasible for pathologists to perform such a quantitative analysis on whole slide images and across multiple cases due to the amount of effort and time required for this purpose. As a consequence, computational approaches for patch-level cellular composition analysis in WSIs are needed. However, computational prediction of the presence and quantification of different cell types in a WSI patch of H\&E stained tissue sections can be challenging due to inter-class phenotype similarities and intra-class variations. Moreover, presence of overlapping cells and diffused background complicates the prediction problem further. 

Existing methods for cellular composition prediction can be divided into two major classes: segmentation-based, and regression-based methods. Segmentation-based methods first segment different type of cells in a patch and then obtain cellular counts through post-processing. For example, watershed transform has been used for red and white blood cell segmentation and counting from microscopy images \cite{monteiro2017detecting}. Similar methods have been used for single cell segmentation in microscopy images of budding yeast colonies using immersion simulation based self-organizing transform \cite{wang2016novel}. Segmentation-based methods require accurate cellular boundaries annotation, or in some cases, scribbled annotations \cite{lee2020scribble2label,koohbanani2020nuclick} during training which can be laborious, error-prone and time consuming . Moreover, state-of-the-art cell segmentation methods such as HoVer-Net  \cite{hovernet}, fail to learn a compact patch-level representation from images which can be useful for downstream prediction tasks in CPath.  

To overcome the pixel-level annotation bottleneck, several regression-based methods have been proposed for cellular composition prediction. Regression-based methods use patch-level cell-type counts as target labels for cellular composition prediction. For instance, recently a convolutional neural network (CNN) has been proposed for counting cells in microscopic images \cite{lavitt2021deep}. Similarly, Cohen et al. proposed a deep neural network called Count-ception, which uses overlapping stride for cell counting \cite{paul2017countception}. Beside this, density maps \cite{lempitsky2010learning}, and manifold learning \cite{wang2018manifold} have been used for counting embryonic, and bacterial cells from light microscopy images. Recently, a manifold regularized network was proposed for counting and localizing cells in histology and microscopy images \cite{zheng2020manifold}. However, these methods are restricted to predict the counts of a single cell type having uniform shape.

In this paper, we propose a multi-headed CNN-based method that uses patch and sub-patch level cellular count during training, and predicts the presence and counts of different type of cells in a patch. The main contributions of the proposed work are listed below:

\begin{itemize}
\item The proposed method harnesses patch and sub-patch level cellular count information during training to predict the presence and counts of different type of cells in a patch. 
 \item The proposed method explicitly models rotation invariance in its design through a contrastive learning inspired learning mechanism.
 \item We have assessed the generalization performance of the proposed method on an independent set and demonstrated the transferability  of the learned feature representation for cellular composition prediction over a dataset with different cell types and annotation structure in comparison to the dataset used for training ALBRT.
 \item The proposed method outperforms state of the art segmentation-based method for cellular composition prediction.
 \item We demonstrate that the learned cellular representation is compact, rotation invariant, and can capture inter- and intra-class variations. This representation can be useful for other downstream prediction tasks in CPath.
 \item The source code and dataset used in this work along with a webserver for patch-level cellular composition analysis are available at the URL: \href{https://github.com/engrodawood/ALBRT}{https://github.com/engrodawood/ALBRT}. 
\end{itemize}

\begin{figure*}
    \begin{center}
    \includegraphics[width=0.9\textwidth, height=3.3in]{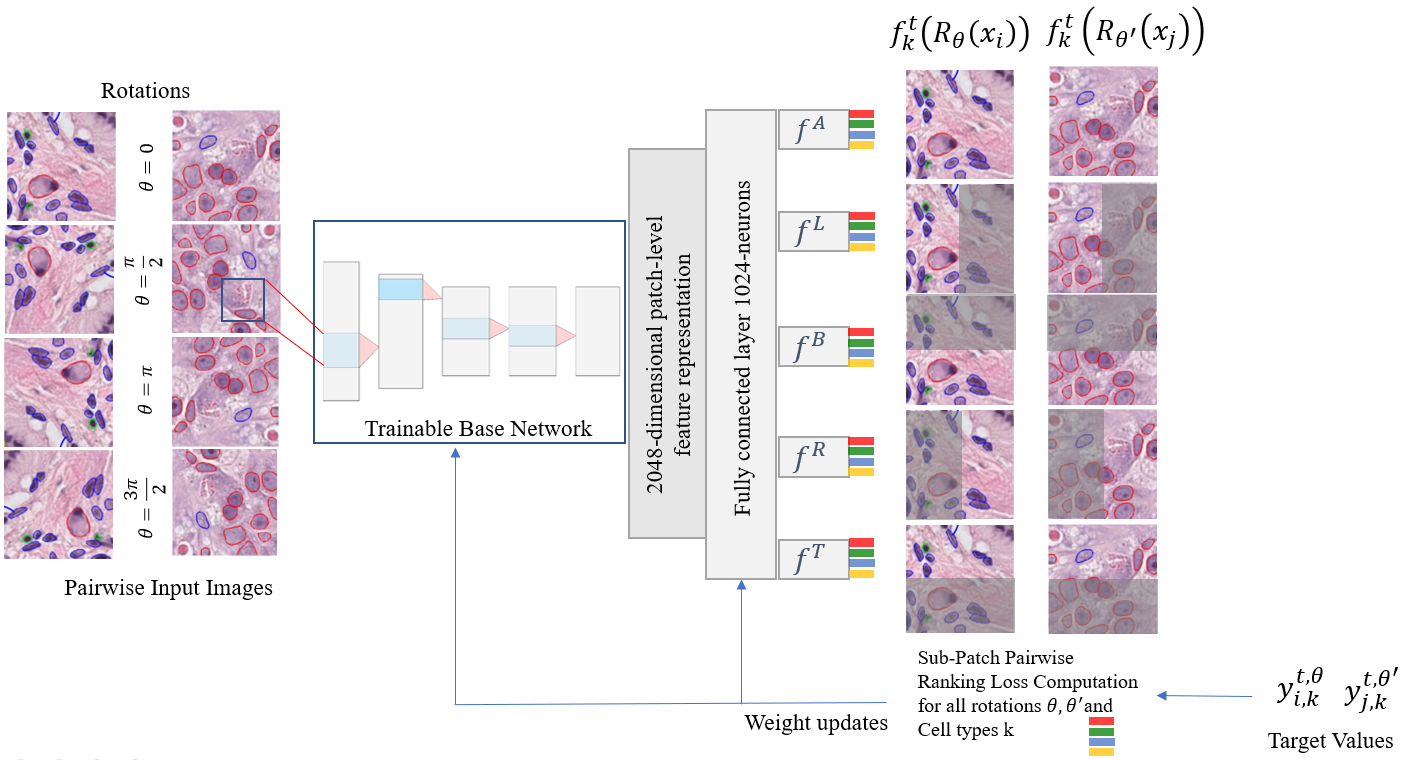}
    \caption{Workflow of the proposed approach for cellular composition prediction.  The model accepts original images along with their rotated variants and predicts cellular counts in different region of the input image. The model has 5-multi-output branches. The first branch (\textit{A}) predicts the counts of different type of cells in the whole of the input patch, whereas the remaining branches (\textit{L,B,R,T}) predict the cellular counts in the left, bottom, right, and top halves of the input patch, respectively. }
    \label{fig:concept}
    \end{center}
\end{figure*}


\section{Materials and Methods}
The workflow of the proposed approach for patch level cellular composition prediction is shown in Figure \ref{fig:concept}. The proposed model takes a patch of size 256×256 pixels from H\&E WSI as input and predicts its cellular composition, i.e., counts of different type of cells in the input patch. The counts of different type of cells predicted by the model are subsequently used for detecting presence or absence of a particular cell type in a patch. Below we provide details about the datasets used in the study, preprocessing, and the proposed model for cellular composition prediction.

\subsection{Datasets and Preprocessing}
We used two publicly available datasets: PanNuke \cite{pannuke,pannukearxiv} and NuCLS \cite{amgad2021nucls}. For model training and validation, PanNuke breast cancer data of four major cell types (Neoplastic, Inflammatory, Connective and Epithelial cells) available in three pre-defined folds was used. The dataset consisting of 2, 351 patches of size 256×256 pixels at a spatial resolution of 0.25 microns-per-pixel (MPP). For each patch, nuclear segmentation masks outlining cellular boundaries of different type of cells in a patch are available. In total, there are 50, 063 annotated nuclei (19, 900 neoplastic, 6, 160 inflammatory, 10, 266 connective and 13, 737 epithelial cells). We compute patch and sub-patch level cellular counts from the nuclear segmentation masks and used those as target labels for cellular composition prediction.  

For assessing model generalization, we used NuCLS dataset~\cite{amgad2021nucls} consisting of 1, 744 field of views (FOVs) partitioned into 5 pre-defined train and validation folds. FOVs from the corrected single-rater NuCLS dataset are of varying sizes with a spatial resolution of 0.20 MPP. To match the resolution of these patches with PanNuke, we extracted a fixed patch of size 320×320 pixels from each FOV and resized it to 256×256 pixels. Patches smaller than 320×320 pixels are excluded. The dataset provides bounding box annotation of various types of nuclei in a patch. The annotated nuclei were grouped into 3 super classes (Tumor, Stromal, and sTILs) and 6 sub-classes (Tumor, Mitotic Figure, Stromal, Macrophages, Lymphocytes, and plasma cells). The pre-processed dataset consists of 33, 205 annotated nuclei, which include 12, 581 Tumor, 145 Mitotic Figure, 5, 949 Stromal, 756 Macrophages, 9, 850 Lymphocytes, and 3, 924 plasma cells.  We calculated patch-level cellular counts of sub-class and super-class cell type by counting cells that lie in a fixed 320×320 window and used the counts as target labels for model performance evaluation. 

For both PanNuke and NuCLS datasets, cells with the majority of their annotated nuclei outside the image patch are excluded from the target count. In addition to cellular counts for each patch, patch-level binary labels indicating the presence or absence of different cell types are also constructed, i.e., if the count of a specific cell type in a patch is greater than zero, the patch would be labelled positive for that cell type, otherwise, negative. These binary cell type labels are used to analyze the sensitivity of various predictors to correctly detect the presence of different types of cells in a given patch. 

\subsection{Proposed Model for Cellular Composition Prediction}
To model cellular composition prediction as a learning problem, consider a training dataset  $D=\{(x_{i},y_{i})| i=1…N\}$ consisting of $N$ patches and their associated cellular counts. Here, $x_{i}$ represents an RGB image patch and $y_{i}\in \mathcal{R^\textit{K}}$ represents the corresponding vector of cellular counts of  $\textit{K}$ different cell types, e.g., neoplastic, inflammatory, connective, and epithelial cells. The objective of the learning task is to train a predictor with a set of learnable parameters $W$ such that the output of the predictor function $z_{i}=f(x_{i};W)$ for $\textit{K}$ cell types matches their true counts for test images. The predictor function was subsequently also used as a classification score for detecting the presence or absence of a particular cell type $\textit{k}$ in a patch.  

We propose a mutli-headed convolutional neural network that uses patch and sub-patch level cellular counts for image based cellular composition prediction, as shown in Figure \ref{fig:concept}. Our hypothesis is that learning sub-patch level features and invariances can lead to precise prediction at the patch level. The proposed model builds on this hypothesis and consists of five multi-output branches with each branch predicting cellular composition within a specific region of the input image patch. The first branch (\textit{A}) predicts the counts of different type of cells in the whole of the input patch, whereas the remaining branches (\textit{L,B,R,T}) predict cellular counts in the left, bottom, right, and top halves of the input patch, respectively. During training, patch and sub-patch level counts of different type of cells in each image are compared with other images and also with their rotated $(90\degree, 180\degree, 270\degree)$ variants through a pairwise ranking-based loss. The proposed model uses Xception network \cite{chollet2017xception} pretrained on ImageNet \cite{ILSVRC15} as a base convolutional network and the flattened output of its convolutional layers is passed onto a fully connected layer consisting of 1024 neurons followed by the five branches for predicting cellular counts in different regions of the input patch. The key motivation behind selecting the Xception network as a base network is that it uses depthwise separable convolution. Since raw patches contain cells of different sizes, depthwise separable convolution with its multiple kernels is expected to capture cell size diversity.

In order to formulate the learning problem, let $f_k^t(R_{\theta}(x_i))$ be the predicted cell counts of branch $t \in \{A,L,B,R,T\}$ for cell type \textit{k} (\textit{k=1, .., K}) for an input patch $x_i$ after it has been rotated by an angle $\theta \in \{0, \frac{\pi}{2}, \pi, \frac{3\pi}{2}\}$. Let $y_{ik}^{t \theta}$ be the corresponding target counts. We use pairwise ranking loss function $l_k()$ as in Equation \ref{l_k}, which compares the predicted cellular composition of images in corresponding branches with their true counts across multiple rotations and sub-patches. The overall loss function ${L(f;D)}$ in Equation \ref{l_F} performs pairwise comparison of all patches in the training dataset across all cell types in a batch. For efficient implementation, the computation of the loss function is vectorized at the batch level. 

\begin{equation}
l_{k}(f;(x_{i},y_{ik}),(x_{j},y_{jk})) = \sum_{t\in \mathcal \textsuperscript{\{A,L,B,R,T\}}}\sum_{\theta}\sum_{\theta '}
\begin{cases}
  \max(0,1-(y_{ik}^{t\theta}-y_{jk}^{t\theta '})(f_{k}^t(R_{\theta}(x_{i})-f_{k}^t(R_{\theta '}(x_{j}))) &  if  y_{ik}^{t\theta} \neq y_{jk}^{t\theta ' }\\
  {(f_{k}^t(R_{\theta}(x_{i})-f_{k}^t(R_{\theta '}(x_{j}))^{2}}& otherwise 
\end{cases}
\label{l_k}
\end{equation}
\begin{equation}
  L (f;D) = \sum_{k=1}^{K}\sum_{i=1}^{N}\sum_{j=1}^{N}
  l_{k}(f;(x_{i},y_{ik}),(x_{j},y_{jk}))
  \label{l_F}
\end{equation}

Minimization of this loss function penalizes the number of mis-ranked examples and attempts to rank patches and sub-patches correctly based on their relative cellular counts. In line with the concept of contrastive learning, it enables the network to learn rotational invariance and sub-patch level translation invariance. Due to the use of a pairwise ranking loss, the predicted cellular counts of the network will be relative and are converted into absolute counts by simple curve-fitting over predicted counts using the ground truth training data.    

\subsection{Model training and evaluation}
We evaluated the performance of the proposed model using 3-fold cross-validation with pre-defined folds in the PanNuke dataset, i.e., data of one-fold was held out for testing and training was performed on the remaining two folds. We assessed the model's regression performance on the test fold using mean absolute error (MAE) and Spearman correlation (SCC) as performance metrics. In order to analyze the model sensitivity in detecting presence of a particular cell type, the predicted count of that cell type was used as prediction score and compared with the corresponding actual binary label (presence or absence) for that patch using the Area under the Receiver Operating Characteristic Curve (AUROC).  In each cross validation run, the proposed model was trained for 15 epochs using adaptive momentum-based optimizer \cite{kingma2014adam} with a batch size of 32 and initial learning rate of 0.001. However, the learning rate by reduced by a factor of 10 after five successive epochs. The experiments were performed using Keras \cite{chollet2015keras} and TensorFlow \cite{tensorflow2015-whitepaper} deep learning libraries.

\subsection{Baseline Method and Comparison}
We compared the performance of proposed method with the state-of-the-art  nuclear segmentation and classification method HoVer-Net \cite{hovernet}. HoVer-Net is a multi-headed CNN that uses set of residual units as backbone feature extractor followed by three branches for cells segmentation and classification. We assessed the performance of HoVer-Net on PanNuke dataset and used their results as baseline for the proposed method. 

\subsection{Transferability  of Cellular Feature Representation}
In order to assess whether the 2048-dimensional patch-level feature representation (see Figure \ref{fig:concept}) learned by the proposed model using the PanNuke dataset can be employed for predicting cellular counts for a different dataset with novel cell types and annotation structure, we used the NuCLS dataset which contains cellular count information for multiple cell-types divided into super and sub-classes. Since cell types in NuCLS dataset are different from PanNuke, we finetuned only the fully connected layers of ALBRT main branch (A) on super-class and sub-class cellular counts. Furthermore, to assess the ability of the learned feature representation to inherently model rotational invariance, no additional augmentations were used. As a baseline for comparison, we used the Xception model pretrained over ImageNet \cite{ILSVRC15}  with both mean-squared error and pairwise ranking losses without any rotational augmentations.

\section{Results and discussion}
\subsection{Quantitative results}
\begin{figure}
    \begin{center}
    \includegraphics[width=0.9\textwidth,height=2.2in]{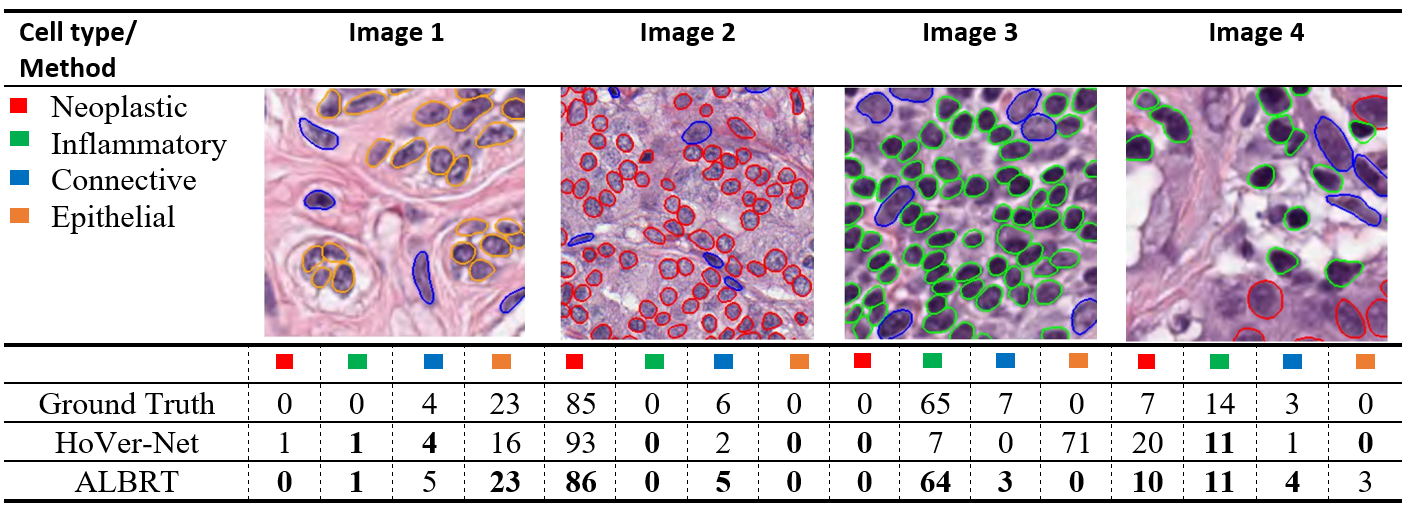}
    \caption{Comparison of ALBRT and HoVer-Net prediction with ground truth for a set of test images. Images are overlaid with annotated cellular boundaries. Below each image true cellular counts, HoVer-Net and ALBRT predicted cellular counts for different types of cells are shown.  }
    \label{fig:test_prediction}
    \end{center}
\end{figure}

Table \ref{table: pannuke comparisions} shows the  performance of the proposed approach for cellular composition prediction in terms of mean MAE and spearman correlation coefficient. From the table, it can be seen that the counts of all cell types listed in the table are predicted with an average MAE of less than 1.78, and a Spearman correlation higher than 0.77. Moreover, for counting epithelial and inflammatory cells, the MAE is around 1.08, while for neoplastic, and connective cell the MAE is 1.78 and 1.71, respectively. Similarly, neoplastic, inflammatory, connective and epithelial cells are predicted with a higher Spearman correlation of 0.94, 0.77, 0.85 and 0.93, respectively. Figure \ref{fig:test_prediction} shows model predictions for a set of patches. The figure visualizes input patches with cellular boundaries outlined using different colors along with their true and predicted cellular counts. From the figure, it can be seen that the difference between the true counts and predicted counts is minimal for all test images.

\begin{table}
    \caption{HoVer-Net and ALBRT cellular composition prediction and classification performance comparison for different types of cells present in PanNuke dataset. We report mean absolute error (MAE), Spearman Correlation coefficient (SCC) and AUROC. The numbers in parenthesis show the standard deviation across test folds.\\}
    \centering
    \input{tables/table_pannuke}
    \label{table: pannuke comparisions}
\end{table}

\subsection{Comparison with HoVer-Net}
Table \ref{table: pannuke comparisions} shows HoVer-Net prediction performance on PanNuke dataset in terms of mean MAE and Spearman correlation coefficient. From the table, it can be seen that for all cell types the MAE of HoVer-Net is higher in comparison to ALBRT. Similarly, the Spearman correlation of ALBRT for all cell types is also higher than HoVer-Net. It is important to mention here that ALBRT uses only cellular count information and does not need precise cellular boundaries annotation during training which can be difficult and time consuming to obtain. 
\subsection{Saliency Map Analysis}
\begin{figure}
    \begin{center}
    \includegraphics[width=0.90\textwidth, height=2.4in]{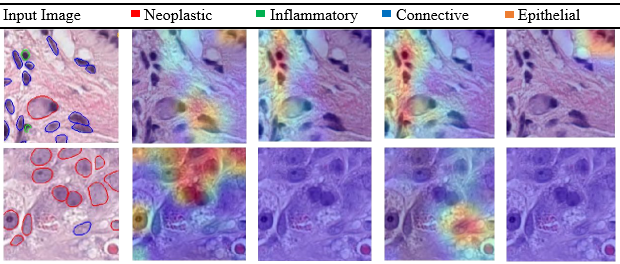}
    \caption{Saliency maps for two text patches. The first column shows the input images overlayed with annotated cellular boundaries in different colors. The other four columns show the saliency map of each cell type obtained using Grad-CAM }
    \label{fig:saliency_maps}
    \end{center}
\end{figure}

We computed gradient-based saliency maps to visualize regions of an input patch that contribute to cellular composition  prediction. We passed set of test images as input to a pretrained model and cellular counts of each cell type as a target and extracted the saliency maps using  Grad-CAM \cite{selvaraju2017grad} as shown in Figure \ref{fig:saliency_maps}. From the figure, it can be seen that for each cell type the extracted saliency map correlates with input patch regions where a specific type of cell is present. This clearly shows that the representation learned by ALBRT truly captures the information that contributes to cellular composition. 

\subsection{Cell Type presence prediction }
\begin{figure}
\vspace{-0.5cm}
\begin{center}
         \subfigure[]{\includegraphics[width = 0.40\textwidth]{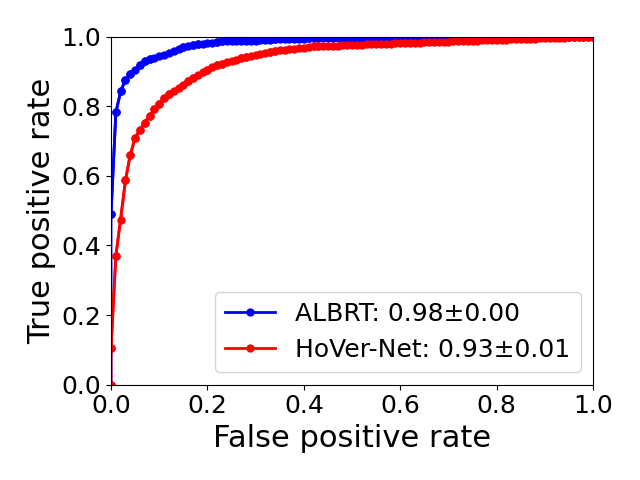}}
         \subfigure[]{\includegraphics[width = 0.40\textwidth]{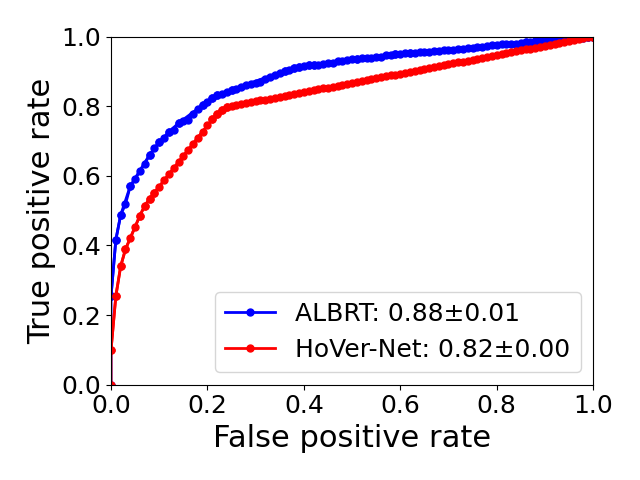}}\\
         \vspace{-0.3cm}
          \subfigure[]{\includegraphics[width = 0.40\textwidth]{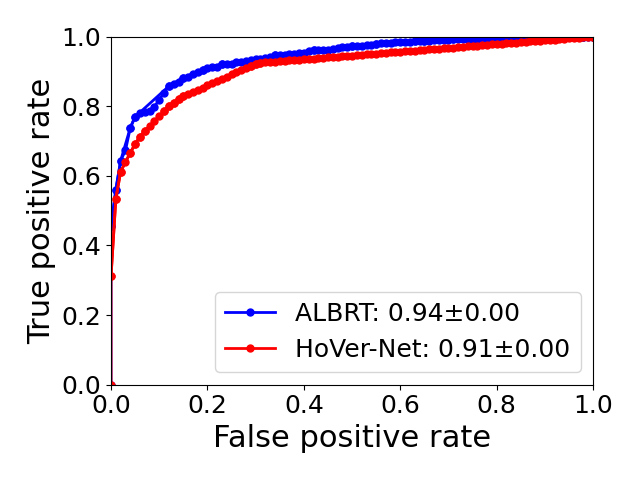}}
          \subfigure[]{\includegraphics[width = 0.40\textwidth]{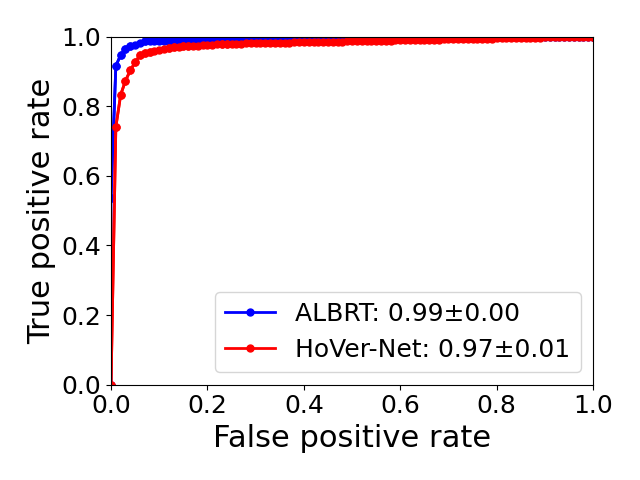}}\\
        \caption{AUROC Comparisons of HoVer-Net and ALBRT for different types of cell present in PanNuke dataset. (a) Neoplastic (b) Inflammatory (c) Connective (d) Epithelial}
    \label{fig:AUROC}
\end{center}
\end{figure}

 Figure \ref{fig:AUROC} shows the AUROC comparison of HoVer-Net and ALBRT. From the figure, it can be seen that the AUROC of ALBRT is significantly higher compared to HoVer-Net. This clearly shows that the proposed approach is capable of detecting the presence of a particular type of cell in a patch and can be used for both detecting the presence of different types of cells as well as counting cells of each type in the patch. 

\subsection{ALBRT captures interclass and intraclass cellular variability }
An interesting feature of ALBRT is that it captures intra-class phenotype similarities and inter-class phenotype variations. To illustrate this, we used the 1024-dimensional output of the final fully connected layer in the network for a given patch as a latent representation  and fit an unsupervised Uniform Manifold Approximation and Projection (UMAP) \cite{mcinnes2018umap} as shown in Figure 2. From the figure, it can be seen that the learned latent representation is capable of classifying cells into different classes. Additionally, it can also cluster patches based on their cell count similarity in a given class. This clearly shows that the latent representation learned by ALBRT truly contribute to cellular composition and cell type classification. Moreover, the representation can be used for other downstream classification and regression problems in CPath. 

\begin{figure}
    \begin{center}
    \includegraphics[width=0.9\textwidth]{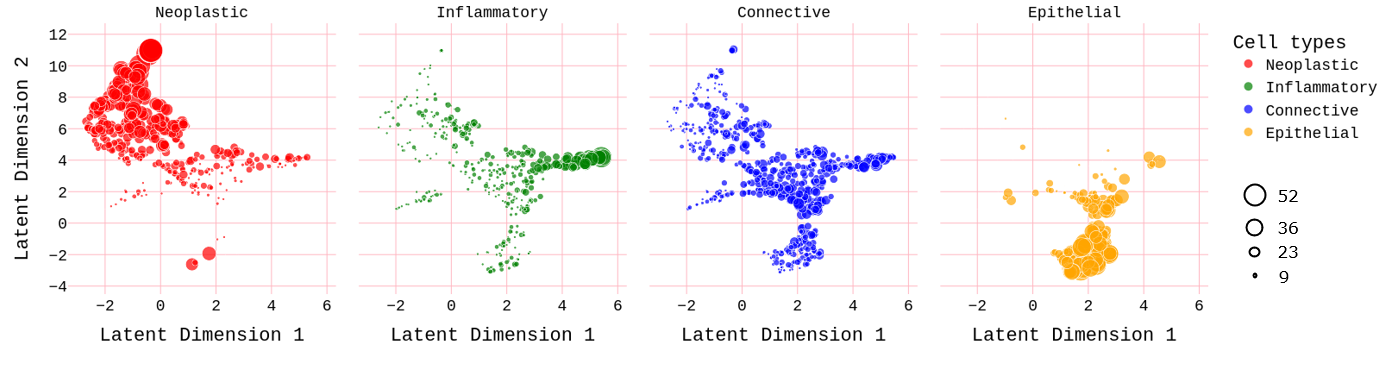}
    \caption{UMAP plots: each blob represents the latent representation of a given input image. The plot on the left shows the latent representation of Neoplastic cells, followed by inflammatory cells, and then connective cells and the final one shows representation of Epithelial cells.}
    \label{fig:umap}
    \end{center}
\end{figure}

\begin{table}
    \begin{center}
    \caption{Results of cellular composition prediction using an Xception model with different patch-level feature representations over NuCLS dataset super classes in terms of mean absolute error (MAE), Spearman Correlation coefficient (SCC), and AUROC. The numbers in parenthesis show the standard deviation across test folds. The patch-level feature representation produced by the Xception network trained under the proposed ALBRT framework over the PanNuke dataset exhibits better prediction performance over the NuCLS dataset in comparison to baseline methods indicating transferability  of the learned feature representation. 
    \\}
    \centering
    \input{tables/superclass}
    \label{table: superclass}
    \end{center}
\end{table}

\begin{table}
    \begin{center}
    \caption{Results of cellular composition prediction with different patch-level feature representations over NuCLS dataset sub-classes in terms of mean absolute error (MAE), Spearman Correlation coefficient (SCC), and AUROC. The numbers in parenthesis show the standard deviation across test folds. The patch-level feature representation produced by the proposed ALBRT framework trained over the PanNuke dataset exhibits better prediction performance for the NuCLS dataset in comparison to baseline methods indicating transferability  of the learned feature representation to novel datasets.\\}
    \centering
    \input{tables/sub-class-types}
    \label{table: subclass}
    \end{center}
\end{table}

\subsection{Transferability  of ALBRT's Cellular Feature Representation}
Tables \ref{table: superclass} and \ref{table: subclass} show ALBRT's generalization performance comparison with ALBRT using ImageNet weights for base network, and Xception network pretrained on ImageNet for super-class, and sub-class cell types. From the tables, it can be seen that for both super-class and sub-class cell types, ALBRT-based feature representation outperforms the baseline Xception model (with ImageNet pre-training and MSE or pairwise-ranking losses) in all performance metrics. For instance, ALBRT-based feature representation predicts the cellular counts of Tumor and sTILs with a mean Spearman correlation of 0.79, and 0.85, whereas ImageNet-based features give a correlation of 0.64, and 0.75. This shows that the feature representation learned by ALBRT is transferable for cellular composition prediction on other datasets. Furthermore, ALBRT using pretrained ImageNet weights for base network outperforms Xception network that uses MSE loss for finetuning. This shows that the loss function used for model training is a good choice for cellular composition prediction compared to MSE.

\section{Conclusions and Future work}
In this work, we have proposed an image-based cellular composition predictor that accurately predicts the presence and counts of different types of cells in a given patch. Overcoming the shortcoming of segmentation-based cellular composition prediction approaches of using precise pixel-level cellular boundaries annotation, we proposed a ranking-based model that learns compact, and rotation-invariant representation from patches using patch and sub-patch level cellular counts information during training. We have assessed the transferability  of the learned representation using the proposed method on an external dataset with novel cell types and annotation structure. Our analysis show that the learned representation is transferrable to other dataset and can be used for various downstream prediction tasks in CPath. In future, we plan to do more detailed comparative analysis on any other cellular composition datasets as well as extending the approach for downstream predictive tasks such as survival analysis. 

\section*{Acknowledgments}
MD would like to acknowledge the PhD studentship support from GlaxoSmithKline. FM and NR supported by the PathLAKE digital pathology consortium which is funded from the Data to Early Diagnosis and Precision Medicine strand of the government’s Industrial Strategy Challenge Fund, managed and delivered by UK Research and Innovation (UKRI).

\bibliographystyle{unsrt}  

\end{document}

%% file: tables/table_pannuke.tex
    \begin{tabu} to \textwidth {XXXXXX}
        \toprule
        Method & Metric & Neoplastic &	Inflammatory &	Connective & Epithelial \\
        \midrule
        HoVer-Net & MAE & 2.34 (0.17) & 1.16 (0.06) & 1.88 (0.04) & 1.55 (0.25) \\
        & SCC &	0.89 (0.02) & 0.72 (0.00)	& 0.79 (0.03) & 0.84 (0.48) \\
        & AUROC & 0.93 (0.01) & 0.82 (0.00) & 0.91 (0.00) & 0.97 (0.01) \\
        
        \midrule
        ALBRT & MAE & \textbf{ 1.78 (0.09)}  & \textbf{1.08 (0.03)} & \textbf{1.71 (0.04)} & \textbf{1.08 (0.09)}\\
         &  SCC & \textbf{0.94 (0.01)} & \textbf{0.77 (0.01)} &  \textbf{0.85 (0.01)} & \textbf{0.93 (0.02)} \\
          & AUROC & \textbf{0.98 (0.00)} & \textbf{0.88 (0.01)} & \textbf{0.94 (0.00)} & \textbf{0.99 (0.00)}\\
          \bottomrule
    \end{tabu}

%% file: tables/superclass.tex
 \begin{tabu} to \linewidth { X[4,L] X[0.78,L] X[1.1,L] X[1.1,L] X[1.1,L]}
    \toprule
    Method & Metric & Tumor &	Stromal &	sTIls \\
   
   \midrule 
    \multirow{3}{=}{ImageNet based features with MSE Loss (Baseline)} & MAE & 5.17 (0.47) & 3.95 (0.43) & 5.37 (0.34) \\
	        & SCC & 0.66 (0.02) & 0.50 (0.04)	& 0.69 (0.07) \\
	        &AUROC	& 0.84 (0.02) & 0.76 (0.04) & 0.81 (0.04) \\
	\midrule
	 \multirow{3}{=}{ImageNet based features with Pairwise-Ranking Loss} & MAE & 4.95 (0.72) & 3.65 (0.48) & 4.38 (0.71) \\
	               & SCC & 0.64 (0.03) & 0.51 (0.02) & 0.75 (0.05) \\
	               & AUROC & 0.84 (0.03) & 0.77 (0.03) & 0.84 (0.03) \\
	
	\midrule
    \multirow{3}{=}{ALBRT features with Pairwise-Ranking Loss}	& MAE & \textbf{3.51 (0.65)} & \textbf{3.15 (0.37)} & \textbf{3.54 (0.79)} \\
    	    & SCC & \textbf{0.79 (0.05)} & \textbf{0.64 (0.05)} & \textbf{0.85 (0.05)}\\
    	    & AUROC	& \textbf{0.91 (0.03)} & \textbf{0.83 (0.04)} & \textbf{0.92 (0.03)}\\
    \bottomrule
\end{tabu}

%% file: tables/sub-class-types.tex
\begin{tabu} to \linewidth { X[2.0,L] X[0.76,L] X[1.2,L] X[1.2,L] X[1.2,L] X[1.2,L] X[1.2,L] X[1.3,L]}
    \toprule
     Method & Metric & Tumor & Mitotic &	Stromal & Macrophage & Lymphocyte &	Plasma cells \\ 
     \midrule
     \multirow{3}{=}{ImageNet based features with MSE Loss (Baseline)} & MAE & 5.05 (0.46) & 0.26 (0.01) & 3.39 (0.45) & 1.17 (0.18) & 5.43 (0.87) & 5.06 (0.74) \\
     & SCC & 0.67 (0.02) & 0.09 (0.07) & 0.54 (0.06) & 0.22 (0.05) & 0.58 (0.07) & 0.21 (0.12) \\
     & AUROC & 0.84 (0.03) & 0.63 (0.11) & 0.77 (0.05) & 0.71 (0.06) & 0.77 (0.04) & 0.63 (0.08)\\
     \midrule
      \multirow{3}{=}{ImageNet based features with Pairwise-Ranking Loss}   & MAE & 4.90 (0.76) & 0.16 (0.09) & 3.14 (0.57) & 0.88 (0.17) & 4.83 (1.3) & 3.67 (1.25)  \\
     & SCC &  0.64 (0.03) & 0.02 (0.04) & 0.55 (0.02) & 0.16 (0.02) &  0.62 (0.07) &  0.28 (0.08) \\
     & AUROC & 0.85 (0.03) & 0.54 (0.07) & 0.78 (0.02) & 0.65 (0.03) & 0.79 (0.03) &  0.68 (0.05) \\\\
     \midrule
      \multirow{3}{=}{ALBRT features with Pairwise-Ranking Loss} & MAE & \textbf{3.47 (0.64)} & \textbf{0.15 (0.08)} & \textbf{2.74 (0.48)}    & \textbf{0.78 (0.18)}  & \textbf{3.81 (1.11)}  & \textbf{3.02 (1.05)}\\
        & SCC & \textbf{0.79 (0.05)} & \textbf{0.16 (0.04)} & \textbf{0.67 (0.04)} & \textbf{0.33 (0.07)} & \textbf{0.74 (0.07)} & \textbf{0.51 (0.03)}\\
        & AUROC & \textbf{0.91 (0.04)} &  \textbf{0.70 (0.04)} & \textbf{0.84 (0.03)} & \textbf{0.80 (0.06)} & \textbf{0.88 (0.04)} & \textbf{0.83 (0.02) }\\\\
 \bottomrule
\end{tabu}